\documentclass[10pt]{article}
\usepackage[T1]{fontenc}
\usepackage{authblk}
\usepackage{amsmath,amssymb,amsfonts}
\usepackage{graphicx}
\usepackage{caption,subcaption}
\usepackage[margin=3cm]{geometry} 
\usepackage[onehalfspacing]{setspace}
\usepackage{cite}
\usepackage{hyperref}
\hypersetup{
    colorlinks=true,
    linktoc=page,
    linkcolor=blue,     
    urlcolor=blue,
    citecolor=blue,
    }
\begin{document}
\title{\boldmath Symmetry Resolved Entanglement Entropy in Hyperbolic de Sitter Space}
\author{Himanshu Gaur\thanks{194123018@iitb.ac.in} }
\author{Urjit A. Yajnik\thanks{yajnik@iitb.ac.in}}
\affil{\textit{Department of Physics, Indian Institute of Technology Bombay, Powai, Mumbai, Maharashtra 400076 India}}
\date{}
\maketitle
\begin{abstract}
In this paper, we study the relation between entanglement and global internal symmetries on de Sitter space. We consider two symmetric causally disconnected regions in the hyperbolic chart on de Sitter space. Since entanglement measures characterises correlations, the study of entanglement between the two causally disconnected regions gives information about the long range correlations in de Sitter space. When a theory possesses an additive global internal symmetry, the entanglement measures for a state with fixed global charge may be decomposed into local charge sectors in either subsystem and thus providing a finer resolution of entanglement. Here we will consider two theories: free complex scalar field, and free Dirac field on de Sitter space. Both theories possess global internal $U(1)$ symmetry. We study the symmetry resolved entanglement entropy for both theories in the Bunch-Davies vacuum state. We find that the symmetry resolved entanglement entropy has equipartition into local charge sectors upto the terms that scale as $V_{H_3}^0$ in the limit of large $V_{H_3}$, where $V_{H_3}$ is the volume of either region. This equipartition however is only broken by the terms of order $O(1/V_{H_3})$. Consequently, we have equipartition of symmetry resolved entanglement entropy in the limit of infinite volume.
\end{abstract}
\clearpage

\tableofcontents
\section{Introduction}
The phenomenon of entanglement is one of the most fascinating feature of quantum theories. Study of entanglement has provided key insights in quantum many body systems \cite{amico2008entanglement,calabrese2004entanglement,vidal2003entanglement}, black hole thermodynamics \cite{solodukhin2011entanglement} and gauge gravity duality \cite{ryu2006holographic,ryu2006aspects}.

The study of long range correlations in de Sitter space is particularly enlightening. Due to the expansion in de Sitter space, the mutually disjoint observers eventually become causally disconnected, but however still remain correlated. The long range correlations thus contain information about the causally disconnected degrees of freedom. To this end study of entanglement between two causally disconnected regions of de Sitter space is of considerable interest, since entanglement characterizes correlations. In the seminal work of Ref. \cite{maldacena2013entanglement}, the entanglement entropy between the two symmetric causally disconnected regions in hyperbolic chart on de Sitter space was studied. They considered a scalar field with minimal coupling in Bunch-Davies vacuum and showed that the entanglement entropy is non vanishing. Since then there have been several investigations of entanglement entropy \cite{kanno2017vacuum, kanno2014entanglement, iizuka2016entanglement, bhattacharya2019emergent, choudhury2019quantum, akhtar2020open, bhattacharya2020dirac, fischler2014entanglement, kanno2015cosmological,albrecht2018quantum,Arias:2019pzy,Arenas-Henriquez:2022pyh}, entanglement negativity \cite{kanno2015entanglement,kukita2017entanglement}, quantum discord \cite{kanno2016quantum,wu2022gaussian,bhattacharya2020some}, and bell type inequalities \cite{maldacena2016model, kanno2017infinite, choudhury2017bell, choudhury2018entangled, bhattacharya2020some,feng2018bell} on de Sitter space in various contexts.

To study entanglement we partition the system into two parts, namely $A$ and its compliment $B$, such that the Hilbert space has the decomposition $\mathcal{H}=\mathcal{H}_{A}\otimes\mathcal{H}_{B}$. If the system is in state $|\Psi\rangle$, the entanglement entropy $S$ is given by
\begin{equation}
S=-\mathrm{Tr}\left[\rho_A\ln\rho_A\right],
\end{equation}
where the reduced density matrix $\rho_A$ is given by $\mathrm{Tr}_B|\Psi\rangle\langle\Psi|$. We now consider a theory that possesses an internal global symmetry such that the corresponding conserved charge is additive i.e. $Q=Q_A+Q_B$, where $Q$ is conserved global charge and $Q_{A}$, and $Q_{B}$ are local charge in region $A$, and $B$ respectively. In such theories the entanglement entropy may be decomposed into local charge sectors \cite{goldstein2018symmetry}. The symmetry decomposition of entanglement has been shown to have equipartition in lower dimensional critical systems with global $U(1)$ symmetry \cite{bonsignori2019symmetry, fraenkel2020symmetry, murciano2020entanglement}. Symmetry decomposition of entanglement has also been studied in the context of $\text{AdS}_3/\text{CFT}_2$ correspondence \cite{zhao2021symmetry, weisenberger2021symmetry, zhao2022charged,belin2013holographic,milekhin2021charge}.

In the present work, we study the symmetry decomposition of entanglement entropy between two causally disconnected regions in de Sitter space into the local charge sectors for theories with global $U(1)$ symmetry. We will consider two theories, namely minimally coupled complex scalar field, and minimally coupled Dirac field. Both the theories possess an internal global $U(1)$ symmetry and are considered to be in the Bunch-Davies vacuum state. We will work in the hyperbolic chart on de Sitter space. In this chart we naturally obtain two symmetrical causally disconnected regions namely $R$, and $L$. The study of symmetry resolved entanglement provides a finer characterisation of long range correlations in de Sitter space.

This paper is organised as follows. In Section \ref{sec2} we briefly discuss the hyperbolic chart on de Sitter space. In Section \ref{sec3} we discuss the symmetry decomposition of entanglement entropy in theories with global $U(1)$ symmetry. In Section \ref{sec4} we find the symmetry resolved entanglement entropy for complex scalar field. In Section \ref{sec5} we find the symmetry resolved entanglement entropy for Dirac field. Finally, in Section \ref{sec6} we discuss our results and conclude this work. 
\section{Hyperbolic de Sitter space} \label{sec2}
In this section we will discuss the hyperbolic chart on de Sitter space. We consider the Euclidean de Sitter space and prepare hyperbolic chart on the Lorentzian de Sitter space by analytically continuing a global chart on Euclidean de Sitter space. This is done in order to find the Bunch Davies vacuum or Euclidean vacuum as discussed in Ref. \cite{sasaki1995euclidean}. The hyperbolic de Sitter consists of three open regions $R$, $L$, and $C$.

The Euclidean de Sitter may be considered as a hypersphere of radius $H^{-1}$ (Hubble constant of de Sitter) in a 5-dimensional Euclidean space. The hypersphere is given by parametrising the Euclidean co-ordinates $(\tilde{x}^0,x^1,x^2,x^3,x^4)$
\begin{equation}
\tilde{x}^0=\cos\tau\cos r,\;x^1=\sin\tau,\; 
\left(
\begin{array}{c}
x^2\\
x^3\\
x^4
\end{array}
\right)=\cos\tau\sin r
\left(
\begin{array}{c}
\cos\theta\\
\sin\theta\cos\varphi\\
\sin\theta\sin\varphi
\end{array}
\right),
\end{equation}
where $(-\pi/2\leq\tau\leq \pi/2)$ and $(0\leq r\leq\pi)$. The metric on the hypersphere is given by
\begin{equation}
\mathrm{d}s_E^2=H^{-2}\left(\mathrm{d}\tau^2+\cos^2\tau\left(\mathrm{d}r^2+\sin^2r\mathrm{d}\Omega^2\right)\right).
\end{equation}
By analytically continuing $\tilde{x}^0\to x^0$, we obtain a chart on the Lorentzian de Sitter space. This chart on de Sitter may be divided into three patches $R$, $L$, and $C$ and they are described by the co-ordinates
\begin{align}
&\left\{
\begin{array}{ll}
t_R=i(\tau-\pi/2),& (t_R\geq 0)\\
r_R=i r, & (r_R\geq 0)
\end{array}
\right.\\
&\left\{
\begin{array}{ll}
t_C=\tau, & (\pi/2\geq t_C\geq -\pi/2)\\
r_C=i(r-\pi/2), & (\infty \geq r_C\geq -\infty)
\end{array}
\right.\\
&\left\{
\begin{array}{ll}
t_L=i(-\tau-\pi/2),& (t_L\geq 0)\\
r_L=i r, & (r_L\geq 0)
\end{array}
\right.,
\end{align}
with their respective metrics
\begin{align}
\mathrm{d}s_R^2&=H^{-2}\left(-\mathrm{d}t_R^2+\sinh^2t_R\left(\mathrm{d}r_R^2+\sinh^2r_R\mathrm{d}\Omega^2\right)\right)\\
\mathrm{d}s_C^2&=H^{-2}\left(\mathrm{d}t_C^2+\cos^2t_R\left(-\mathrm{d}r_C^2+\cos^2r_C\mathrm{d}\Omega^2\right)\right)\\
\mathrm{d}s_L^2&=H^{-2}\left(-\mathrm{d}t_L^2+\sinh^2t_L\left(\mathrm{d}r_L^2+\sinh^2r_L\mathrm{d}\Omega^2\right)\right).
\end{align}
\begin{figure}
\centering 
\includegraphics[width=0.35\textwidth]{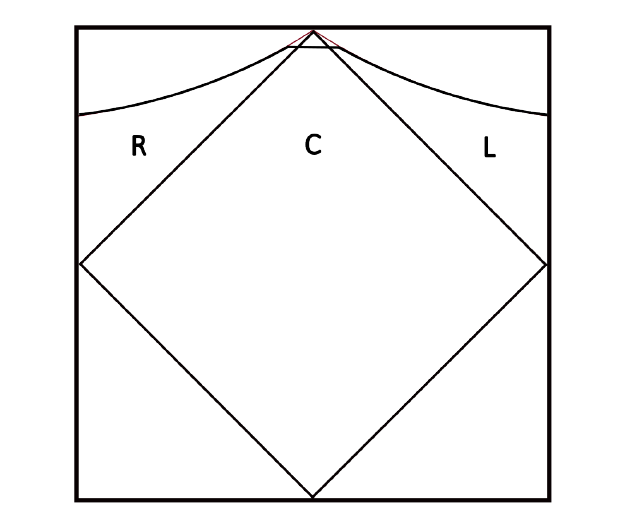}
\caption{\label{fig:i} Penrose diagram for hyperbolic chart on de Sitter space. The three patches $R$, $L$, and $C$ are denoted in the figure. Thinner red line on either $R$ and $L$ patch are constant $t_R$ and $t_L$ surfaces respectively. The thick black line is the chosen Cauchy surface.}
\end{figure}
The Penrose diagram for the hyperbolic chart on de Sitter is shown in Figure \ref{fig:i}. Now in a field theory to find the global euclidean modes a Cauchy surface must be chosen. This is done by choosing an large hyperbolic radius $r_R$ and $r_L$ on either patch at $t_R=\text{constant}$ and $t_L=\text{constant}$ respectively. Finally, these two parts are connected via a bridge through $C$. This construction is shown in Figure \ref{fig:i}. When the radius on either patch is appropriately large, we may ignore the bridge while normalising the modes on the Cauchy surface.
\section{Symmetry resolution of entanglement} \label{sec3}
In this section we will discuss the symmetry decomposition of entanglement into the local charge sectors for a system with global $U(1)$ symmetry \cite{goldstein2018symmetry}. We will consider that our system is partitioned into two parts, namely $A$ and its compliment $B$. The partition is such that the Hilbert space has the decomposition $\mathcal{H}=\mathcal{H}_A\otimes \mathcal{H}_B$.

We further assume that the charge $\hat{Q}$ corresponding to the global $U(1)$ symmetry is additive in $A$ and $B$ i.e. $\hat{Q}=\hat{Q}_A+\hat{Q}_B$. When the system is in a state $|\Psi\rangle$ of fixed global charge we have the relation $[\rho,\hat{Q}]=0$, where $\rho=|\Psi\rangle\langle\Psi |$. Taking the trace over $B$ degrees of freedom gives $[\rho_A,\hat{Q}_A]=0$. This implies that the RDM $\rho_A$ is block diagonal in local charge sectors of $\hat{Q}_A$. We may project $\rho_A$ into the local charge sector $q$ using the projection operators $\Pi_q$
\begin{equation} \label{eq3.1}
p_q\rho_{A,q}=\Pi_q\rho_A\Pi_q,
\end{equation}
where $q$ is an eigenvalue of $\hat{Q}_A$, $\rho_{A,q}$ is the density matrix in charge sector $q$ and $p_q$ is the probability of getting $q$ upon measurement of $\hat{Q}_A$. The study of entanglement in the charge sectors $q$ has been termed symmetry resolved entanglement. The R\'enyi entropy and entanglement entropy in the local charge sector $q$ is given by
\begin{align}
S_{n,q}&=\frac{1}{1-n}\mathrm{Tr}\ln\left[\rho_{A,q}^n\right], \label{eq3.2}\\
S_{q}&=-\partial_n\mathrm{Tr}\ln\left.\left[\rho_{A,q}^n\right]\right|_{n\to 1} \label{eq3.3}
\end{align}
respectively. This block structure of density matrix however may not manifest easily and for this purpose charged moments $Z_n(\alpha)$ are introduced. For the present case the charged moments are given by
\begin{equation}\label{eq3.4}
Z_n(\alpha)=\mathrm{Tr}\left[\rho_A^n e^{i\alpha\hat{Q}_A}\right].
\end{equation}
Taking their fourier transform gives the trace of $\left(p_q\rho_{A,q} \right)^n$. For brevity we denote this via $\mathcal{Z}_n(q)$ and it is given by
\begin{equation}\label{eq3.5}
\mathcal{Z}_n(q)=\frac{1}{2\pi}\int_{-\pi}^{\pi}\mathrm{d}\alpha\, Z_n(\alpha)e^{-iq\alpha}.
\end{equation}
The symmetry resolved R\'enyi entropy $S_{n,q}$, and Entanglement entropy $S_{q}$ in charge sector $q$ are then given by using eq.\eqref{eq3.5} in eq.\eqref{eq3.2} and eq.\eqref{eq3.3} respectively,
\begin{align}
S_{n,q}&=\frac{1}{1-n}\ln\left[\frac{\mathcal{Z}_n(q)}{\mathcal{Z}_1^n(q)}\right], \label{eq3.6}\\
S_{q}&=-\partial_n\ln\left.\left[\frac{\mathcal{Z}_n(q)}{\mathcal{Z}_1^n(q)}\right]\right|_{n\to 1}. \label{eq3.7}
\end{align}
\section{Complex scalar field} \label{sec4}
In this section we find the symmetry resolved entanglement entropy of complex scalar field on hyperbolic de Sitter space.

We consider a complex scalar field with a minimal coupling on the de Sitter space, whose action reads
\begin{equation} \label{eq4.1}
S_S=-\int\mathrm{d}^4x\,\sqrt{-g}\,\left(g^{\mu\nu}\nabla_\mu \phi^{*}\nabla_\nu \phi +m^2\phi^{*}\phi\right).
\end{equation}
Complex boson field has a global $U(1)$ symmetry under the transformation $\phi\to\phi\,e^{i\alpha}$ and $\phi^{*}\to\phi^{*} e^{-i\alpha}$, with the corresponding conserved charge $Q$
\begin{equation} \label{eq4.2}
Q=i\int\mathrm{d}S^{\mu}\sqrt{-g}(\phi\,\partial_{\mu}\phi^{*}-\phi^{*}\partial_{\mu}\phi) 
\end{equation}
where $\mathrm{d}S^{\mu}$ is the differential surface element on the Cauchy surface and $i\sqrt{-g}(\phi\,\partial_{\mu}\phi^{*}-\phi^{*}\partial_{\mu}\phi)$ is the conserved current $J_{\mu}$. As discussed in Section \ref{sec2} in hyperbolic slicing of de Sitter space we have three patches. Here we are interested in the correlations between the disconnected patches $R$ and $L$. Since, for the Cauchy surface chosen in Section \ref{sec2} we may ignore the $C$ patch, what we then have is effectively a theory on $R\cup L$. The mode equation on either patch is given by
\begin{equation} \label{eq4.3}
\left[\frac{1}{\sinh^3t}\frac{\partial}{\partial t}\sinh^3t\frac{\partial}{\partial t}-\frac{1}{\sinh^2t}\boldsymbol{L}^2+\frac{9}{4}-\nu^2 \right]u(t,r,\Omega)=0,
\end{equation}
where $\boldsymbol{L}^2$ is the Laplacian on the unit 3-hyperboloid and the parameter $\nu$ introduced above is
\begin{equation}
\nu=\sqrt{\frac{9}{4}-\frac{m^2}{H^2}}. \label{eq4.4}
\end{equation}
The case of coupling $\xi$ to the scalar curvature may be similarly considered after replacing $m^2$ by $m_{eff}^2=m^2+12\xi H^2$ in eq.(\ref{eq4.4}). We have the case of conformally coupled massless scalar at $\nu=\frac{1}{2}$ and minimally coupled massless scalar at $\nu=\frac{3}{2}$. The corresponding local charge $Q_A$ on the $R$ patch is given by
\begin{equation} \label{eq4.5}
Q_A=\int_R\mathrm{d}^3\,x\,J_0,
\end{equation}
where the integral is over the constant $t_R$ surface.
\subsection{Symmetry resolved entanglement}
The global Bunch Davies vacuum however does not co-inside with the $R$, and $L$ vacua. This is easily seen by taking the mode expansion of the field $\phi$ in terms of globally well defined modes and locally well defined modes on either patch \cite{sasaki1995euclidean}. The global vacuum state for boson field when written in $R$ and $L$ modes is a Gaussian state. The exact expression for scalar field has been obtained in the Ref. \cite{maldacena2013entanglement}. The case of complex boson can be handled similarly and the Bunch Davies vacuum in terms of the positive frequency mode operator $a_{i,p,l,m}$, and negative frequency mode operator $b_{i,p,l,m}$ on the $R$ and $L$ patch is given by
\begin{equation} \label{eq4.6}
|{\Omega}\rangle_{p,l,m}\propto e^{m_{ij}a_i^{\dagger}b_j^{\dagger}}|{R}\rangle |{L}\rangle,
\end{equation}
where $a_{R}|R\rangle=b_{R}|R\rangle=0$, $a_{L}|L\rangle=b_{L}|L\rangle=0$, $i\in\left\{R,L\right\}$ and
\begin{equation} \label{eq4.7}
m_{ij}=\frac{\sqrt{2} e^{-p \pi}}{\sqrt{\cosh 2 \pi p+\cos 2 \pi \nu}}\left(\begin{array}{cc}
\cos \pi \nu & i \sinh p \pi \\
i \sinh p \pi & \cos \pi \nu
\end{array}\right).
\end{equation} 
The mode operators satisfy the commutation relations $[a_{i,p,l,m},a_{i',p',l',m'}^{\dagger}]$ $=[b_{i,p,l,m},b_{i',p',l',m'}^{\dagger}]$ $=\delta_{i,i'}\delta_{p,p'}\delta_{l,l'}\delta_{m,m'}$, and rest are zero. In eq.\eqref{eq4.6} we suppressed the quantum numbers $(p,l,m)$ on operators $a_i$, and $b_i$ and will continue to do so in what follows for brevity. The quantum numbers $(p,l,m)$ corresponds to the Laplacian on the hyperbola appearing in eq.\eqref{eq4.3}. The local charge operator in terms of bosonic mode operators is
\begin{equation} \label{eq4.8}
Q_A=\sum_{p,l,m}a^{\dagger}_{R}a_{R}-b^{\dagger}_{R}b_{R}.
\end{equation}
We now introduce the Schmidt basis so that the vacuum state may be written as
\begin{equation} \label{eq4.9}
|\Omega\rangle_{p,l,m} \propto e^{\gamma_p\left(\tilde{a}_{R}^{\dagger}\tilde{b}_{L}^{\dagger}+\tilde{b}_{R}^{\dagger}\tilde{a}_{L}^{\dagger}\right)}|\tilde{R}\rangle |\tilde{L}\rangle,
\end{equation}
with the relation $\tilde{a}_{R}|\tilde{R}\rangle=\tilde{b}_{R}|\tilde{R}\rangle=0$ and $\tilde{a}_{L}|\tilde{L}\rangle=\tilde{b}_{L}|\tilde{L}\rangle=0$. The two sets of Bosonic operators are related via the Bogoliubov transformation 
\begin{align}
\tilde{a}_{R}&=\alpha {a}_{R}+ \beta {b}_{R}^{\dagger}, \qquad \tilde{b}_{R}=\alpha {b}_{R}+ \beta {a}_{R}^{\dagger}\label{eq4.10}\\
 \tilde{a}_{L}&={\alpha}^* {a}_{L}+ {\beta}^* {b}_{L}^{\dagger}, \qquad \tilde{b}_{L}={\alpha}^* {b}_{L}+ {\beta}^* {a}_{L}^{\dagger}.\label{eq4.11}
\end{align}
The mode operators satisfy the commutation relation $[\tilde{a}_i,\tilde{a}_{i'}^{\dagger}]=[\tilde{b}_i,\tilde{b}_{i'}^{\dagger}]=\delta_{i,i'}$ and consequently the Bogoliubov coefficients satisfy the relation $|\alpha|^2-|\beta|^2=1$. From eq.\eqref{eq4.6}, eq.\eqref{eq4.7}, and eq.\eqref{eq4.9}-eq.\eqref{eq4.11} we obtain $\gamma_p$ (in eq.\eqref{eq4.9}) to be
\begin{equation} \label{eq4.12}
\gamma_p=i \frac{\sqrt{2}}{\sqrt{\cosh 2 \pi p+\cos 2 \pi \nu}+\sqrt{\cosh 2 \pi p+\cos 2 \pi \nu+2}}.
\end{equation}
The reduced density matrix in the Schmidt basis is then obtained by taking the trace over $L$ degrees of freedom. The reduced density matrix is given by
\begin{equation} \label{eq4.13}
\left(\rho_{A}\right)_{p,l,m}=\frac{1}{\left(1-|\gamma_p|^2\right)^2}\sum_{j,k=0}^{\infty}|\gamma_p|^{2j+2k}|j,k,p,l,m\rangle \langle j,k,p,l,m|.
\end{equation}
The local charge operator $\hat{Q}$ in terms of the new bosonic operators is easily obtained from eq.\eqref{eq4.10} and eq.\eqref{eq4.11} to be
\begin{equation} \label{eq4.14}
\hat{Q}_A=\sum_{p,l,m}\tilde{a}_{{R}}^\dagger \tilde{a}_{{R}}-\tilde{b}_{{R}}^\dagger \tilde{b}_{{R}}.
\end{equation}
The charged moments in sector $(p,l,m)$ are obtained using the eq.\eqref{eq4.13} and eq.\eqref{eq4.14} in eq.\eqref{eq3.4} and are given by
\begin{equation} \label{eq4.15}
\left(Z_n\left(\alpha\right)\right)_{p,l,m}=\frac{\left(1-|\gamma_p|^2\right)^{2n}}{1-2|\gamma_p|^{2n}\cos\left(\alpha\right)+|\gamma_p|^{4n}}.
\end{equation}
The summation over all modes $(p,l,m)$ is equivalent to the volume of the hyperbola and an integration in $p$ over the density of states \cite{maldacena2013entanglement}. The charged moments are given by
\begin{equation} \label{eq4.16}
Z_n(\alpha)=\exp\left\{V_{H_3}\int_0^{\infty}\mathrm{d}p\, \frac{p^2}{2\pi^2}\ln\left(\frac{\left(1-|\gamma_p|^2\right)^{2n}}{1-2|\gamma_p|^{2n}\cos\left(\alpha\right)+|\gamma_p|^{4n}}\right) \right\}, 
\end{equation}
where $V_{H_3}$ is the volume of the $R$ patch on the Cauchy surface and $\frac{p^2}{2\pi^2}$ is the density of modes for bosons on the hyperbola \cite{bytsenko1996quantum}. To evaluate $\mathcal{Z}_n(q)$ using eq.\eqref{eq3.5}, we first note that the volume $V_{H_3}\gg 1$ and take the taylor expansion of the exponent in eq.\eqref{eq4.16} upto $O(\alpha^2)$. We have
\begin{align}
\int_0^{\infty}\mathrm{d}p\frac{p^2}{2\pi^2}\ln\left(\frac{\left(1-|\gamma_p|^2\right)^{2n}}{1-2|\gamma_p|^{2n}\cos\left(\alpha\right)+|\gamma_p|^{4n}}\right)&\approx\int_0^{\infty}\mathrm{d}p\frac{p^2}{2\pi^2}\ln\left(\frac{\left(1-|\gamma_p|^2\right)^{2n}}{1-2|\gamma_p|^{2n}+|\gamma_p|^{4n}}\right)-\alpha^2\frac{\Lambda_n^s}{2}, \label{eq4.17}\\
\text{where,}\qquad\Lambda_n^s &=\int_0^{\infty}\mathrm{d}p\frac{p^2}{2\pi^2}\frac{2|\gamma_p|^{2n}}{1-2|\gamma|^{2n}+|\gamma|^{4n}}. \label{eq4.18}
\end{align}
This is a good approximation for $Z_n(\alpha)$ in the range $\alpha\in (-\pi,\pi)$. The fourier transform of $Z_n(\alpha)$ is evaluated by taking the Gaussian approximation of the integral to be
\begin{equation} \label{eq4.19}
\mathcal{Z}_{n}(q)\approx \frac{Z_{n}(0)}{\sqrt{2\pi \Lambda_n^s V_{H_3}}}\exp\left\{-\frac{q^2}{2\Lambda_n^s V_{H_3}}\right\}.
\end{equation}
The quantity $\mathcal{Z}_{1}(q)$ is the charge distribution probability of the local charge $\hat{Q}_A$. We observe that the charge $q$ has a Gaussian distribution with the standard deviation being proportional to the volume $V_{H_3}$. The symmetry resolved entanglement entropy is obtained by using eq.\eqref{eq4.19} in eq.\eqref{eq3.5}
\begin{equation} \label{eq4.20}
S_q\approx S-\frac{1}{2}\ln\left[V_{H_3}\right]+\left\{\frac{1}{2}\left.\frac{\partial_n\Lambda_n^s}{\Lambda_n^s}\right|_{n\to 1}-\frac{1}{2}\ln\left[2\pi\Lambda_1^s\right]\right\}-\frac{q^2}{2V_{H_3}}\left(\left.\frac{\partial_n\Lambda_n^s}{\left(\Lambda_n^s\right)^2}\right|_{n\to 1}+\frac{1}{\Lambda_1^s}\right),
\end{equation}
where $S$ is the entanglement entropy of the complex scalar field. We notice that the symmetry resolved entanglement entropy in the large $V_{H_3}$ limit has an equipartition into the charge sectors $q$ upto the leading order terms in the volume $V_{H_3}$. This equipartition is broken only by the terms of order $O\left(1/V_{H_3}\right)$. Hence, in the limit $V_{H_3}\to \infty$ we have equipartition to all orders.

\begin{figure}
\centering 
\includegraphics[width=1\textwidth]{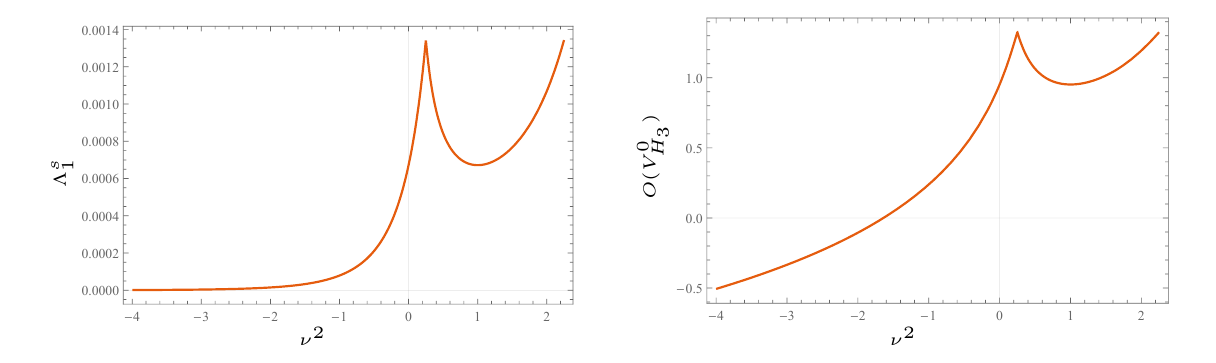}
\caption{\label{fig:ii} On the left is plotted $\Lambda_{1}$ given by eq.\eqref{eq4.18}. The standard deviation for the charge distribution is proportional to $\Lambda_{1}$. It is shown to have a local maxima at $\nu=\frac{1}{2}$ and $\nu=\frac{3}{2}$. On the right is plotted "curly bracket term" in eq.\eqref{eq4.20} and it is denoted by $O\left(V_{H_3}^0\right)$ above. This term is also shown to have a local maxima at $\nu=\frac{1}{2}$ and $\nu=\frac{3}{2}$.}
\end{figure}

We see from Figure \ref{fig:ii} that $\Lambda_1^s$ has a local maxima at conformally coupled massless scalar and minimally coupled massless scalar cases, reflecting higher probability of non-zero local charge at these points than other values of $\nu^2$. The term that scale as $V_{H_3}^0$ (i.e "curly bracket term") in eq.\eqref{eq4.20} is plotted in Figure \ref{fig:ii}. This contribution to symmetry resolved entanglement entropy is again maximised at conformally coupled massless scalar and minimally coupled massless scalar cases, similar to the entanglement entropy behaviour for the scalar field. We also notice the strange behaviour for both the plots in Figure \ref{fig:ii} in the low mass regime $\left(\nu \geq \frac{1}{2}\right)$, a similar behaviour is also present in entanglement entropy of the scalar field \cite{maldacena2013entanglement}, since we would expect correlations to become stronger as mass decreases. We mention here that there is a supercurvature mode $\left(p=i\left(\nu-\frac{1}{2}\right)\right)$ in this regime \cite{sasaki1995euclidean}. The entanglement contribution, if any, of this mode has not been calculated so far and can possibly alter the behaviour of entanglement in this region. This mode is not considered in this work as well. However it is true that this supercurvature mode is not entangled with any other mode since it has a different value of $p$.

\section{Dirac field} \label{sec5}
In this section we find the symmetry resolved entanglement entropy for Dirac field on hyperbolic de Sitter space.

We consider the Dirac field on the de Sitter space with minimal coupling, whose action reads
\begin{equation} \label{eq5.1}
S_D=\int\mathrm{d}^4\,x\sqrt{-g}\bar{\Psi}\left(\gamma^{\mu}\mathrm{D}_\mu -m\right)\Psi.
\end{equation}
Dirac field has a global internal $U(1)$ symmetry under $\Psi\to e^{i\alpha}\Psi$, and $\Psi^{\dagger}\to e^{-i\alpha}\Psi^{\dagger}$ and the corresponding conserved charge is
\begin{equation} \label{eq5.2}
Q=-\int\mathrm{d}\,S_{\mu}\bar{\Psi}\gamma^{\mu}\Psi,
\end{equation}
where $\mathrm{d}\,S_{\mu}$ is the differential surface element on the Cauchy surface. We consider the hyperbolic chart on de Sitter space just as we did in the last section. The problem of computing the Bunch Davis vacuum and entanglement entropy for Dirac field in the same setting has been considered in Ref.\cite{kanno2017vacuum}. The local charge $Q_A$ on the $R$ patch is
\begin{equation} \label{eq5.3}
Q_A=\int_R\mathrm{d}^3\,x\sqrt{-g}\Psi^{\dagger}\Psi,
\end{equation}
where the integral is on the constant $t_R$ surface.
\subsection{Symmetry resolved entanglement}
In the case of fermions we may write the Bunch Davies vacuum $|\Omega\rangle$ as a direct product $|0\rangle^{+}|0\rangle^{-}$. The state $|0\rangle^{+}$ $\left(|0\rangle^{-}\right)$ is annihilated by spin down (spin up) global positive frequency mode, and spin up (spin down) global negative frequency mode annihilation operators. The state $|0\rangle^{+}$ in terms of mode operators on the $R$ patch $c^R_{(\downarrow,p,l,m)}$, $d^R_{(\uparrow,p,l,m)}$ and on the $L$ patch $c^L_{(\uparrow,p,l,m)}$, $d^L_{(\downarrow,p,l,m)}$ is given by
\begin{equation} \label{eq5.4}
|0\rangle_{(p,l,m)}^{+}\propto e^{m_{ij}c^{i\dagger}d^{j\dagger}}|0\rangle_R^{+} |0\rangle_L^{-},
\end{equation}
where we have the relations $c^R_{\downarrow}|0\rangle_R^{+}=d^R_{\uparrow}|0\rangle_R^{+}=0$ and $c^L_{\uparrow}|0\rangle_L^{-}=d^L_{\downarrow}|0\rangle_L^{-}=0$. The state $|0\rangle^{-}$ is written similarly by using the complimentary spin mode operators on both patches. The mode operators satisfy the anti-commutation relation $\{c_{s,p,l,m}^i,c_{s',p',l',m'}^{i'\dagger}\}=\{d_{s,p,l,m}^i,d_{s',p',l',m'}^{i'\dagger}\}=\delta_{i,i'}\delta_{s,s'}\delta_{p,p'}\delta_{l,l'}\delta_{m,m'}$, and rest are zero. From here on we will suppress the spins on mode operators as well. The matrix $m_{ij}$ in this case is given by
\begin{align} 
m_{ij}=&-\frac{B^*}{A^2+1}\left(
\begin{array}{cc}
A&-1\\
1&A
\end{array}
\right), \qquad\text{where}\label{eq5.5}\\
A=\frac{\sinh\frac{\pi m}{H}}{\cosh\pi p},& \qquad B=\frac{e^{-\pi p}\Gamma\left(\frac{1}{2}-ip\right)^2}{\Gamma\left(\frac{1}{2}-ip-i\frac{m}{H}\right)\Gamma\left(\frac{1}{2}-ip+i\frac{m}{H}\right)}.\label{eq5.6}
\end{align} 
The local charge operator $Q_A$ in terms of mode operators on the $R$ patch is given by
\begin{equation} \label{eq5.7}
Q_A=\sum_{s,p,l,m}c^{R\dagger}c^{R}-d^{R\dagger}d^{R},
\end{equation}
where $s$ in the summation stands for spin. The Schmidt basis were introduced for this problem through Bogoliubov transformations 
\begin{align}
\tilde{c}_{R}&=u {c}^{R}+ v {d}^{R{\dagger}}, \qquad \tilde{d}_{R}=u {d}^{R}- v {c}^{R\dagger} \label{eq5.8}\\
 \tilde{c}_{L}&={u}^* {c}^{L}- {v}^* {d}^{L\dagger}, \qquad \tilde{d}_{L}={u}^* {d}^{L}+ {v}^* {c}^{L\dagger}.\label{eq5.9}
\end{align}
in the Ref.\cite{kanno2017vacuum}. The mode operators satisfy the anti-commutation relations $\{\tilde{c}_i,\tilde{c}_{i'}^{\dagger}\}=\{\tilde{d}_i,\tilde{d}_{i'}^{\dagger}\}=\delta_{i,i'}$ and rest are zero. The Bogoliubov coefficients satisfy the relation $|u|^2+|v|^2=1$. The state $|0\rangle^{+}$ in the Schmidt basis is given by
\begin{equation} \label{eq5.10}
|0\rangle^{+}_{p,l,m}\propto e^{\gamma_p\left(\tilde{c}_{R}^{\dagger}\tilde{d}_{L}^{\dagger}+\tilde{d}_{R}^{\dagger}\tilde{c}_{L}^{\dagger}\right)}|0\rangle_{R'}^{+}|0\rangle_{L'}^{-},
\end{equation} 
with the relations $\tilde{c}_{R}|0\rangle_{R'}^{+}=\tilde{d}_{R}|0\rangle_{R'}^{+}=0$ and $\tilde{c}_{L}|0\rangle_{L'}^{+}=\tilde{d}_{L}|0\rangle_{L'}^{+}=0$. The parameter $\gamma_p$ is given by
\begin{equation} \label{5.10i}
\gamma_p=\frac{1}{2m_{12}}\left(m_{11}^2+m_{12}^2-1+\sqrt{\left(m_{11}^2+m_{12}^2-1\right)^2+4m_{12}^2}\right).
\end{equation}
The reduced density matrix is then obtained by taking the trace over the $L$ degrees of freedom. It is given by
\begin{equation} \label{eq5.11}
\begin{split}
&{\rho_A}_{s,p,l,m}=\\
&\hspace{0.2in}\frac{1}{\left(1+|\gamma_p|^2\right)^2}\left(|0\rangle^{+}_{R'}{}_{R'}^+\langle 0|+|\gamma_p|^2|1,0\rangle^{+}_{R'}{}_{R'}^+\langle 1,0|+|\gamma_p|^2|0,1\rangle^{+}_{R'}{}_{R'}^+\langle 0,1|+|\gamma_p|^4|1,1\rangle^{+}_{R'}{}_{R'}^+\langle 1,1|\right)
\end{split}
\end{equation}
The local charge operator $Q_A$ on the $R$ patch is obtained by using eq.\eqref{eq5.8}, and eq.\eqref{eq5.9} in eq.\eqref{eq5.7}  
\begin{equation} \label{eq5.12}
Q_{A}=\sum_{s,p,l,m}\tilde{c}_{R'}^{\dagger}\tilde{c}_{R'}-\tilde{d}_{R'}^{\dagger}\tilde{d}_{R'}.
\end{equation}
\begin{figure}
\centering 
\includegraphics[width=1\textwidth]{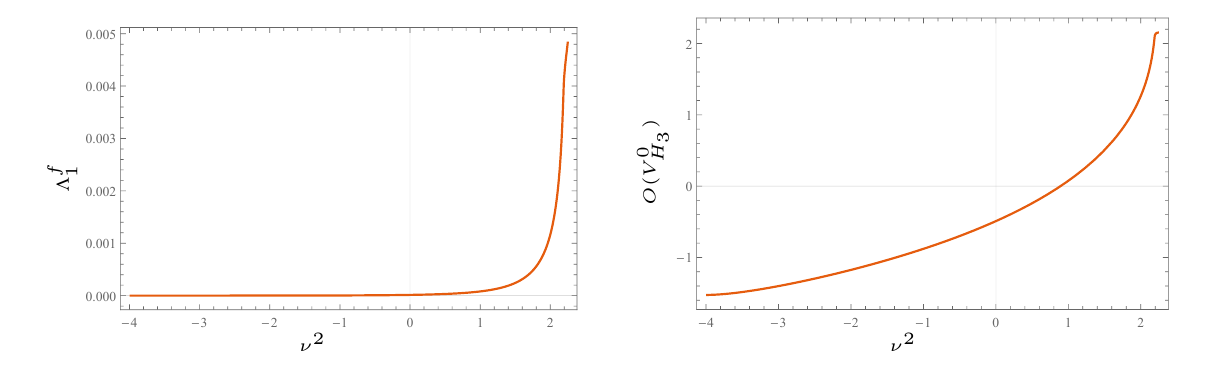}
\caption{\label{fig:iii} On the left is plotted $\Lambda_1^f$ given by eq.\eqref{eq5.14}. The standard deviation of the charge distribution is proportional to $\Lambda_1^f$ and it shown to monotonically increase with $\nu^2$. On the right is plotted the "curly bracket term" in eq.\eqref{eq5.16}. This term is denoted by $O(V_{H_3}^0)$ and is shown to monotonically decrease with $\frac{m^2}{H^2}$.}
\end{figure}
The charged moments for fermions are obtained using the eq.\eqref{eq5.11}, and eq.\eqref{eq5.12} in eq.\eqref{eq3.5}. We have
\begin{align}
Z_{n}(\alpha)&\approx Z_n(0)\exp\left\{-\frac{V_{H_3}\Lambda_n^f\alpha^2}{2}\right\}, \qquad\text{where} \label{eq5.13}\\
\Lambda_n^f&=2\int_0^{\infty}\mathrm{d}p\left(\frac{1/4+p^2}{2\pi^2}\right)|\gamma|^{2n} \label{eq5.14}
\end{align}
here we have only taken terms upto $O(\alpha^2)$ in the exponent and used the density of states for fermions on hyperbola to be $({1/4+p^2})/{2\pi^2}$ \cite{bytsenko1996quantum, camporesi1996eigenfunctions}. $\mathcal{Z}_n(q)$ is obtained by taking the fourier transform of eq.\eqref{eq5.13}. Taking the Gaussian approximation in the integral we obtain
\begin{equation} \label{eq5.15}
\mathcal{Z}_n(q)\approx \frac{Z_n(0)}{\sqrt{2\pi\Lambda_n^f V_{H_3}}}\exp\left\{-\frac{q^2}{2\Lambda_n^f V_{H_3}}\right\}.
\end{equation}
We note that similar to the case of complex scalar field the local charge has a Gaussian distribution with standard deviation proportional to the volume $V_{H_3}$. Using eq.\eqref{eq5.15}, the symmetry resolved entanglement entropy is obtained to be
\begin{equation} \label{eq5.16}
S_q=S-\frac{1}{2}\ln\left[V_{H_3}\right]+\left\{\frac{1}{2}\left.\frac{\partial_n\Lambda_n^f}{\Lambda_n^f}\right|_{n\to 1}-\frac{1}{2}\ln\left[2\pi\Lambda_1^f\right]\right\}-\frac{q^2}{2V_{H_3}}\left(\left.\frac{\partial_n\Lambda_n^f}{\left(\Lambda_n^f\right)^2}\right|_{n\to 1}+\frac{1}{\Lambda_1^f}\right),
\end{equation}
here $S$ is the entanglement entropy for the Dirac field. Similar to the bosonic case we see that the symmetry resolved entanglement entropy has an equipartition into the local charge sectors upto the leading order terms in $V_{H_3}$. The plot for the term that scale as $V_{H_3}^0$ (i.e. "curly bracket term") is shown in Figure \ref{fig:iii}, where it is shown to monotonically increase with $\nu^2=\frac{9}{4}-\frac{m^2}{H^2}$ and hence the contribution to symmetry resolved entanglement entropy at this order monotonically decreases with $\frac{m^2}{H^2}$. The equipartition of entanglement entropy is broken by an $O\left(1/V_{H_3}\right)$ term in this case as well.

In the case of Dirac fermions there is no strange behaviour in the low mass limit. It is also true that there are no supercurvature modes present for Dirac fermions. We also observe that the $O\left(V_{H_3}^0\right)$ contribution to symmetry resolved entanglement entropy is higher in Dirac fermion case than the scalar case in the limit $m\to 0$, similar to the behaviour of entanglement entropy \cite{kanno2017vacuum}.
\section{Conclusion} \label{sec6}
We have studied the symmetry decomposition of entanglement entropy for theories in their ground state with global $U(1)$ symmetry into the local charge sectors on de Sitter background. We considered the entanglement between two causally disconnected regions, namely $R$ and $L$ patch, in the hyperbolic chart on de Sitter space. We studied two cases, free complex scalar field, and free Dirac field. We first showed that the local charge in either region has a Gaussian distribution for both the fields. We then found that for both of these cases the symmetry resolved entanglement entropy has an equipartition into the local charge sectors $q$ upto terms that scales as $V_{H_3}^0$, where $V_{H_3}$ is the volume of the constant time slice of the subregion (i.e. the $R$ patch). We also found that the leading term is just the entanglement entropy for the field in the expression for symmetry resolved entanglement entropy. The equipartition of symmetry resolved entanglement entropy is however broken by a term proportional to the inverse of the same volume. This terms vanishes in the limit of infinite volume and hence we have the equipartition of symmetry resolved entanglement entropy. 

In the low mass regime $\frac{1}{2}\leq\nu\leq\frac{3}{2}$, we saw a strange oscillatory behaviour for the $O\left(V_{H_3}^0\right)$ contribution to the symmetry resolved entanglement entropy in the complex scalar case, similar to the behaviour of the entanglement entropy \cite{maldacena2013entanglement}. In this regime there is a supercurvature mode present whose contribution, if any, was not studied in this work. The contribution from supercurvature mode to entanglement entropy, if any, has not been calculated yet. The $O\left(V_{H_3}^0\right)$ contribution term is also shown to have a local maxima at $\nu=\frac{1}{2}$ (massless conformally coupled field) and $\nu=\frac{3}{2}$ (massless minimally coupled field). In the Dirac field case there are no supercurvature modes and the $O\left(V_{H_3}\right)$ contribution is shown to monotonically decrease with $\frac{m^2}{H^2}$. We also found that this contribution to symmetry resolved entanglement entropy is higher in Dirac field than the scalar field in the limit $m\to 0$.
\section*{Acknowledgments}
HG is supported by the Prime Minister's Research fellowship provided by Ministry of Education, Government of India. UAY is supported by an Institute Chair Professorship.

\bibliographystyle{unsrt}
\bibliography{Biblio.bib}
\end{document}